\documentclass[reprint]{revtex4-2}
\usepackage{lipsum}
\usepackage[load=addn]{siunitx}
\usepackage{xcolor}
\usepackage{graphicx}
\usepackage{amsmath}
\usepackage{braket}
\usepackage[colorlinks]{hyperref}
\hypersetup{
    linkcolor = {blue},
	anchorcolor = {blue},
	citecolor= {blue}
}

\newcommand{\stdNc}{\Delta_\text{c}}

\begin{document}
\title{Stochastic dynamics of a few sodium atoms in a cold potassium cloud}

\author{Rohit Prasad Bhatt}
\author{Jan Kilinc}
\author{Lilo Höcker}
\author{Fred Jendrzejewski}
\affiliation{
Universität Heidelberg, Kirchhoff-Institut für Physik,
Im Neuenheimer Feld 227, 69120 Heidelberg, Germany
}

\date{\today}

\begin{abstract}
We report on the stochastic dynamics of a few sodium atoms immersed in a cold potassium cloud. The studies are realized in a dual-species magneto-optical trap by continuously monitoring the emitted fluorescence of the two atomic species. We investigate the time evolution of sodium and potassium atoms in a unified statistical language and study the detection limits. We resolve the sodium atom dynamics accurately, which provides a fit free analysis. This work paves the path towards precise statistical studies of the dynamical properties of few atoms immersed in complex quantum environments.
\end{abstract}


\maketitle %\maketitle must follow title, authors, abstract and \pacs

\section{Introduction}\label{sec:Intro}

The random evolution of a small system in a large bath can only be described by its statistical properties. Such stochastic dynamics occur in a wide range of settings including financial markets \cite{Cont2000}, biological systems \cite{Kucsko2013}, impurity physics \cite{Grusdt2015} and quantum heat engines \cite{Gluza_2020}.
Their evolution is hard to describe from microscopic principles, stimulating strong efforts to realize highly controlled model systems in optomechanics \cite{Groeblacher2009}, cavity QED \cite{Gleyzes2007}, superconducting circuits \cite{Pekola_2012}, trapped ions \cite{Hauke_2019} and cold atoms \cite{Chiu_2019}. For cold atoms, the high control is complemented by the access to a number of powerful statistical approaches, like the precise analysis of higher-order correlation functions of a many-body system \cite{Schweigler2017} or the extraction of entanglement through fluctuations \cite{Strobel2014, Lukin2019}.

Cold atomic mixtures offer a natural mapping of physical phenomena involving system-bath interactions, wherein one species realizes the bath, while the other species represents the system. If a mesoscopic cloud of the first species is immersed in a Bose-Einstein condensate formed by the second species, it implements the Bose polaron problem \cite{Scelle2013a, Rentrop2016, Jorgensen2016, Yan2020}.
In recent quantum simulators of lattice gauge theories, the small clouds of one species emulate the matter field, which is properly coupled to the gauge field realized by the second atomic species \cite{Zohar2015, Kasper2017, Mil2019}.
Proposed technologies even go towards quantum error correction, where the logical qubits are implemented in one atomic species and the second atomic species mediates entanglement between them~\cite{Kasper_2020}.
The feasibility of immersing a few atoms into a large cloud was demonstrated in a dual-species magneto-optical trap (MOT) of rubidium and cesium \cite{Weber2010}. This was extended towards the study of position- and spin-resolved dynamics with a single tracer atom acting as a probe \cite{Hohmann2017, Bouton2020}.
However, combining such experiments with a statistical description of system and bath remains an open challenge.

\begin{figure}
 \includegraphics[width=1.0 \columnwidth]{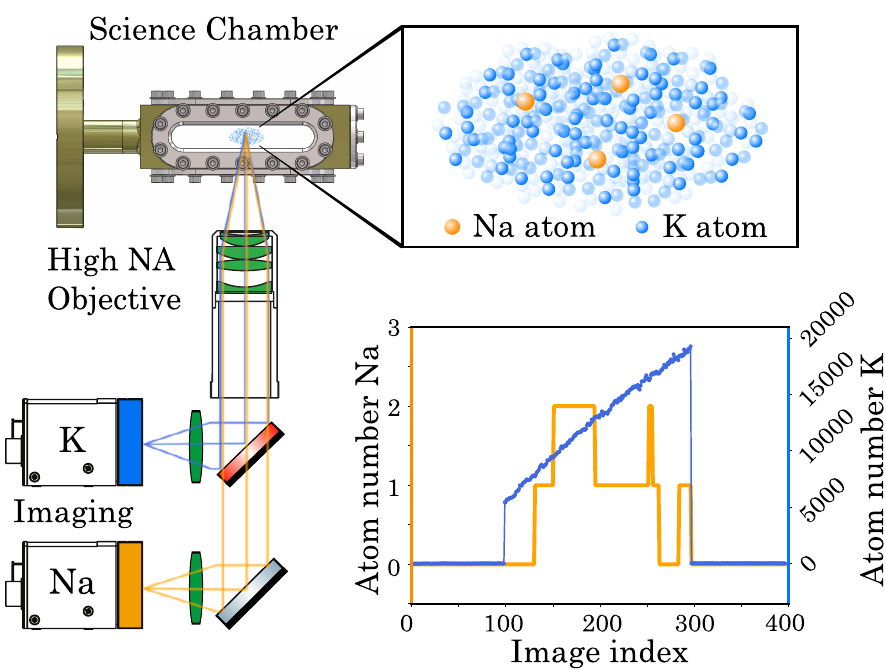}
 \caption{Experimental platform for atom counting. The atoms are trapped and laser cooled in a dual-species MOT inside the science chamber. The emitted fluorescence is collected by a high-resolution imaging system onto the cameras. We observe the stochastic dynamics of single sodium atoms (orange),  immersed in a large cloud of potassium atoms (blue).
\label{fig:MotivationFig}}
\end{figure}

In this work, we investigate the stochastic dynamics of  few sodium atoms and a large cloud of potassium atoms in a dual-species MOT. It builds upon atom counting experiments with a single atomic species \cite{Choi2007,Ueberholz_2002, Wenz2013, Hume2013} and atomic mixtures of Rb and Cs \cite{Weber2010, Widera_2016, Schmidt2016}. The mixture of \textsuperscript{23}Na and \textsuperscript{39}K, as employed in our experiment, has shown excellent scattering properties in the degenerate regime \cite{Schulze2018}. It further provides the option to replace the bosonic \textsuperscript{39}K by the fermionic \textsuperscript{40}K isotope through minimal changes in the laser cooling scheme~\cite{Park2012}. The two atomic species are cooled through standard laser cooling techniques in a dual-species MOT, as shown in figure~\ref{fig:MotivationFig}. In a MOT, cooling and trapping is achieved through a combination of magnetic field gradients with continuous absorption and emission of resonant laser light.
We collect the resulting fluorescence on a dedicated camera for each species and trace their spatially integrated dynamics. We present a statistical analysis for the dynamics of both species, which separates the fluctuations induced by the statistical loading process from those caused by technical limitations. Furthermore, we achieve single atom counting for sodium, which we employ to study its full counting statistics.

The paper is structured as follows. In section~\ref{sec:ExpApp}, we provide a detailed discussion of the experimental apparatus, and how it is designed to fulfill the requirements of modern quantum simulators.
In section~\ref{Sec:dynamics}, we study the dynamics of the observed fluorescence signal for both atomic species. The analysis of their mean and variance after an ensemble average is then employed to statistically investigate the origin of different fluctuations.
In section~\ref{Sec:Immersion}, we leverage the single atom counting resolution of sodium to extract the full counting statistics of atom load and loss events. In section~\ref{Sec:Outlook}, we end with an outlook of the next steps for the experimental platform.

\section{Experimental Apparatus}
\label{sec:ExpApp}
In this section, we describe the different elements of our new experimental apparatus. In the course of designing this machine, effort was taken to optimize the versatility and stability of the system. To achieve this, the experimental setup was designed for a continuous development of the vacuum and the laser system.

\subsection{Vacuum system}
\label{subsec:VacSys}
Ultracold atom experiments require an ultra-high vacuum (UHV) at pressures below $\SI{e-11}{\millibar}$, in order to isolate the cold atoms from the surrounding environment. Our group operates another machine for quantum simulation experiments with atomic mixtures \cite{Scelle2013a, Rentrop2016, Mil2019}, which consists of a dual-species oven and a Zeeman slower connected to a science chamber \cite{Stan2005}. In this apparatus the first cooling stages of the two species are highly coupled, which renders the optimization of the system very complex.

\begin{figure}
 \includegraphics[width=1.0 \columnwidth]{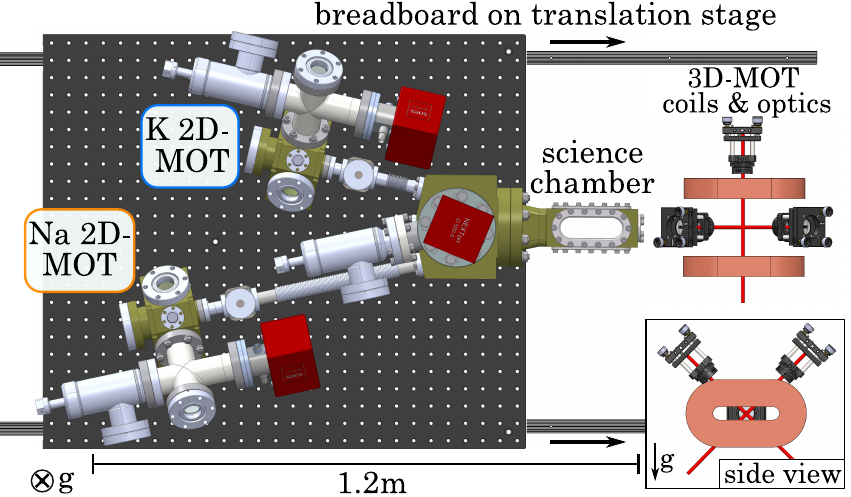}
 \caption{Vacuum system. The separated 2D-MOT chambers are connected from the same side to the dual-species science chamber. The vacuum pumps are shown in red.  The whole vacuum system is mounted on a translation stage, such that the science chamber can be moved out of the region of the 3D-MOT coils and optics.\label{fig:VacuumComplete}}
 \end{figure}

In the new vacuum system, we decoupled the precooling stages of sodium and potassium up to the science chamber, as sketched in figure \ref{fig:VacuumComplete}. The compact vacuum system contains two independent two-dimensional magneto-optical trap (2D-MOT) chambers for sodium and potassium and a dual-species science chamber, where experiments are performed \cite{Tiecke2009, Lamporesi2013a}. The two 2D-MOT chambers are connected to the science chamber from the same side under a $12.5^\circ$ angle. The entire apparatus is mounted on a \SI{600}{\milli\meter} x \SI{700}{\milli\meter} aluminium breadboard, which is fixed to a linear translation stage \footnote{Inspired by the approach of the programmable quantum simulator machine in the Lukin group \cite{Endres2016}.}. Therefore, we are able to move the science chamber out of the contraption of magnetic field coils and optics. This allows for independent improvement of the vacuum system and in-situ characterization of the magnetic field at the position of the atoms.

\textit{2D-MOT.} The design of the 2D-MOT setup is inspired by ref.~\cite{Lamporesi2013a}. The chamber body is manufactured from titanium (fabricated by \textit{SAES Getters}), where optical access is ensured by standard CF40 fused silica viewports with broadband anti-reflection coating (BBR coating). The 2D-MOT region has an oven containing a \SI{1}{\gram} atomic ingot ampoule. The oven is heated to $160\,^\circ$C ($70\,^\circ$C) for sodium (potassium), thereby increasing the pressure to $\SI{e-8}{\millibar}$ in this region. To maintain an UHV in the science chamber, a differential pumping stage separates the two vacuum regions from each other. Two gate valves ensure full decoupling of the two atomic species by isolating different chambers. Each region is pumped with its separate ion getter pump (from \textit{SAES Getters}) \footnote{We use \textit{NEXTorr Z100} for the 2D-MOT and \textit{NEXTorr D500} for the 3D-MOT.}.
We employed four stacks of nine (four) neodymium bar magnets to generate the required magnetic quadrupole field inside the sodium (potassium) 2D-MOT chamber.

\textit{3D-MOT.} The rectangular titanium science chamber is designed such that the two atomic beams from the 2D-MOT chambers intersect in the center. Optical access for various laser beams and a high-resolution imaging system is maximized by four elongated oval viewports (fused silica, BBR coating), which are sealed using indium wire.

The quadrupole magnetic field required for the 3D-MOT is produced by the MOT coils, which are placed on the sides of the science chamber. Applying a current of \SI{20}{\ampere} to the coils results in a magnetic field gradient of $17$\,G/cm. The fast control of the current in the coils, required during an experimental sequence, is achieved through an  insulated-gate bipolar transistor (IGBT) switching circuit. In order to cancel stray fields in the vicinity of the atomic clouds, we use three independent pairs of Helmholtz coils carrying small currents ($< 1$\,A).
\begin{figure*}[t]
	\includegraphics[width=1.0 \textwidth]{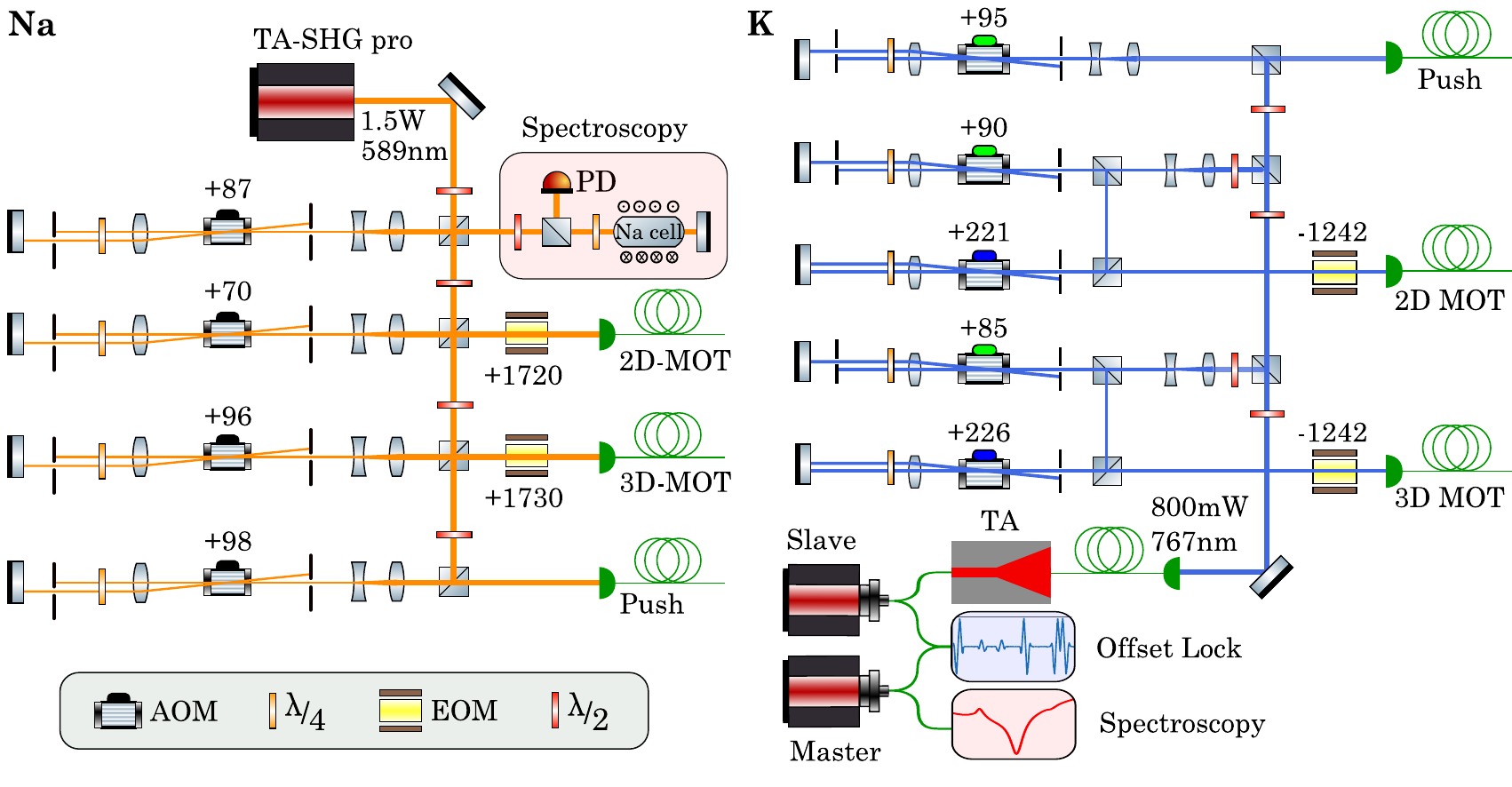}
	\caption{Sketch of the optical setup for laser cooling sodium and potassium atoms. The laser light is split into different paths, enabling the individual control of laser power and frequency for the 2D-MOT, the 3D-MOT, and the push beam. The frequency and intensity of these beams is controlled with the help of acousto-optic modulators (AOMs) in double-pass configuration. The rf-frequencies for AOMs and EOMs are given in MHz. \textbf{Na:} The repumping light for the 2D- and 3D-MOT is generated by electro-optic modulators (EOMs). \textbf{K:} In the 2D- and 3D-MOT paths the green AOM controls the $^{39}$K cooling frequency and the blue AOM is responsible for the creation of the $^{39}$K repumping light. The repumping light for $^{40}$K is generated by EOMs. \label{fig:SodiumLaserTable}}
\end{figure*}

\subsection{Laser cooling}
\label{subsec:LasSys}

In order to cool and trap the atoms, the laser light is amplified and frequency-stabilized on a dedicated optical table for each atomic species. The light is transferred to the main experiment table via optical fibers. The layout of the laser systems for both species is shown in figure~\ref{fig:SodiumLaserTable}.

\textit{Sodium.} Laser cooling and trapping of sodium atoms is achieved using the D$_2$-line at 589 nm, which is obtained from a high-power, frequency-doubled diode laser (\textit{TA-SHG pro}, from \textit{Toptica Photonics}). The laser light is stabilized to excited-state crossover transition of the D$_2$-line using saturated absorption spectroscopy (SAS) and Zeeman modulation locking \cite{Weis_1988}. The modulated SAS signal is fed into a digital lock-in amplifier and PI-controller, which are programmed on a \textit{STEMLab 125-14} board from \textit{Red Pitaya} using the \textit{Pyrpl} module \cite{Neuhaus_2017}.

\textit{Potassium.} Laser cooling and trapping of potassium atoms is achieved using the D$_2$-line at \SI{767}{\nano\meter}. The light is obtained from a master-slave laser configuration (both \textit{DL pro}, from \textit{Toptica Photonics}).  The master laser frequency is locked to the ground-state crossover transition of the D$_2$-line of $^{39}$K with a scheme similar to sodium. The slave laser is frequency-stabilized through an offset beat lock (\SI{405}{\mega\hertz}) and its output is amplified to a power of \SI{800}{\milli\watt}, using a home-built tapered amplifier (TA) module. This light is used to supply all the cooling and trapping beams.
The offset locking scheme also facilitates  switching between the two isotopes,  $^{39}$K and $^{40}$K. To cool the fermionic $^{40}$K, the slave laser frequency is increased by approximately \SI{810}{\mega\hertz} via the offset lock and the blue acousto-optic modulators (see figure~\ref{fig:SodiumLaserTable}) are turned off.

\textit{3D-MOT}. On the experiment table, the light from the optical fibers is distributed into three independent paths for the operation of the dual-species MOT in a retro-reflected configuration. 
For both species the number of atoms loaded into the 3D-MOT can be tuned in a controlled way by adjusting the 2D-MOT beam power and the oven temperature.
The pre-cooled atoms in the 2D-MOT region are transported to the 3D-MOT with a push beam.
For accurate atom counting of sodium, we use \SI{1.3}{\milli\watt} power in each 3D-MOT beam and a beam diameter of about \SI{2}{\milli\meter}, while the push and 2D-MOT beams are turned off. This helps in reducing the loading rate as well as the stray light.
Furthermore, the sodium oven is kept at a relatively low temperature of $80$\,$^\circ$C, which increases the lifetime of atoms in the 3D-MOT due to better vacuum.

\subsection{Fluorescence imaging}
The cold atoms are characterized by collecting their fluorescence through an imaging system with a high numerical aperture (NA) onto a camera (fig.~\ref{fig:MotivationFig}). The imaging setup comprises an apochromatic high-resolution objective, which features an NA of 0.5 and chromatic focal correction in the wavelength range $589-767$\,nm (fabricated by \textit{Special Optics}). The fluorescence of sodium and potassium is separated by a dichroic mirror, built into a cage system, which is mounted on stages for x-, y- and z-translation along with tip-tilt adjustment.

Both imaging paths contain a secondary lens and an additional relay telescope. This allows us to do spatial filtering with an iris in the intermediate image plane of the secondary lens and achieve a magnification of 0.75 (0.25) for sodium (potassium).
For imaging the sodium atoms we use an sCMOS camera (\textit{Andor ZYLA 5.5}) \cite{Picken_2017, Schlederer_2020}, while for the potassium atoms we use an EMCCD camera (\textit{NuVu H-512}).
In total, we estimate the conversion efficiency from photons to camera counts to be $0.2$\% ($0.02$\%) for sodium (potassium).

\section{Atom dynamics}\label{Sec:dynamics}

\begin{figure}
	\includegraphics[width=1.0 \columnwidth]{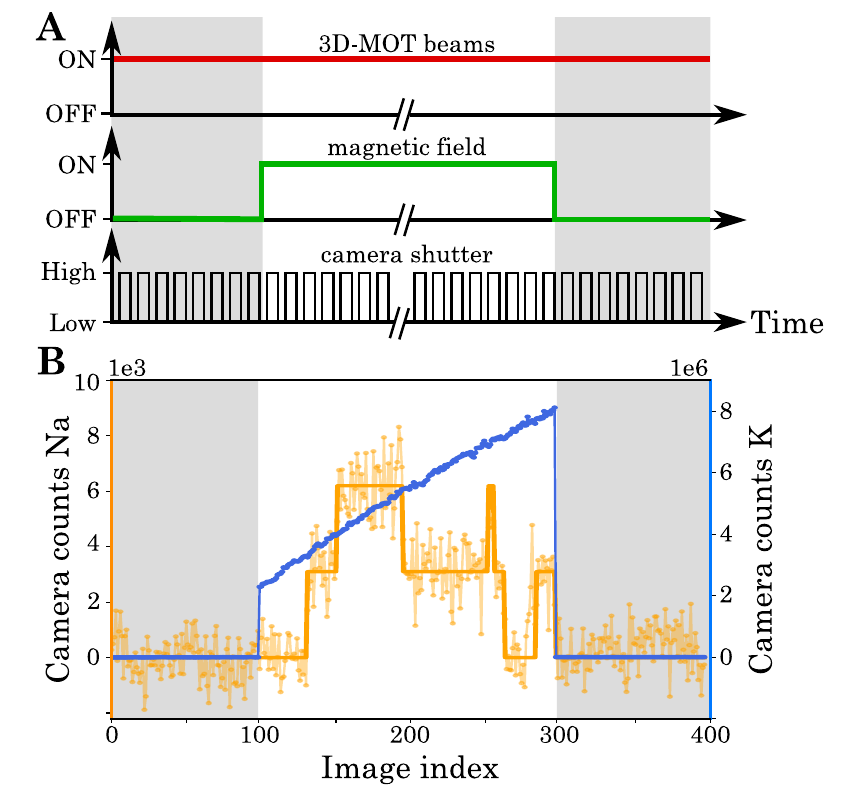}
	\caption{Experimental sequence. \textbf{A:} A series of images (black) is taken. While the MOT beams (red) are always on, the magnetic field (green) is switched off for reference images marked in grey. \textbf{B:} Typical time trace from the series of images of sodium (orange) and potassium (blue).\label{fig:ExSeq}}
\end{figure}

Our experimental sequence to investigate the atom dynamics is shown in figure~\ref{fig:ExSeq}~A.
We start the atom dynamics by switching on the MOT magnetic field (with a gradient of $21$\,G/cm) and then monitor the fluorescence in $N_{\text{img}}=200$ images.
Each image has an integration time $\tau = \SI{75}{\milli\second}$, such that the camera counts overcome the background noise.
Since the motion of the atoms during the integration time washes out any spatial information, we sum up the counts over the entire MOT region for each image.
This results in a time trace of camera counts $N_\text{c}$, as shown in figure \ref{fig:ExSeq}~B.
Each experimental run is preceded and succeeded by a series of 100 reference images to quantify the background noise $\Delta_{\text{bg}}$, induced by the fluctuations in the stray light from the MOT beams.

The camera counts for sodium exhibit random jumps, corresponding to single atom load and loss events. The stochastic nature of the observed signal and large relative fluctuations require a statistical analysis of the dynamics in terms of expectation values.
The single atom resolution provides additional access to the full counting statistics, which is discussed in section~\ref{Sec:Immersion}. The few sodium atoms are immersed in a cloud of potassium atoms, which we pre-load for \SI{5}{\second} to ensure large atom numbers. In contrast to sodium, we do not observe discrete jumps, but rather a continuous loading curve with higher counts and smaller relative fluctuations.
These are typical features of a bath, which can be characterized by its mean and variance.

\begin{figure}
	\includegraphics[width=1.0 \columnwidth]{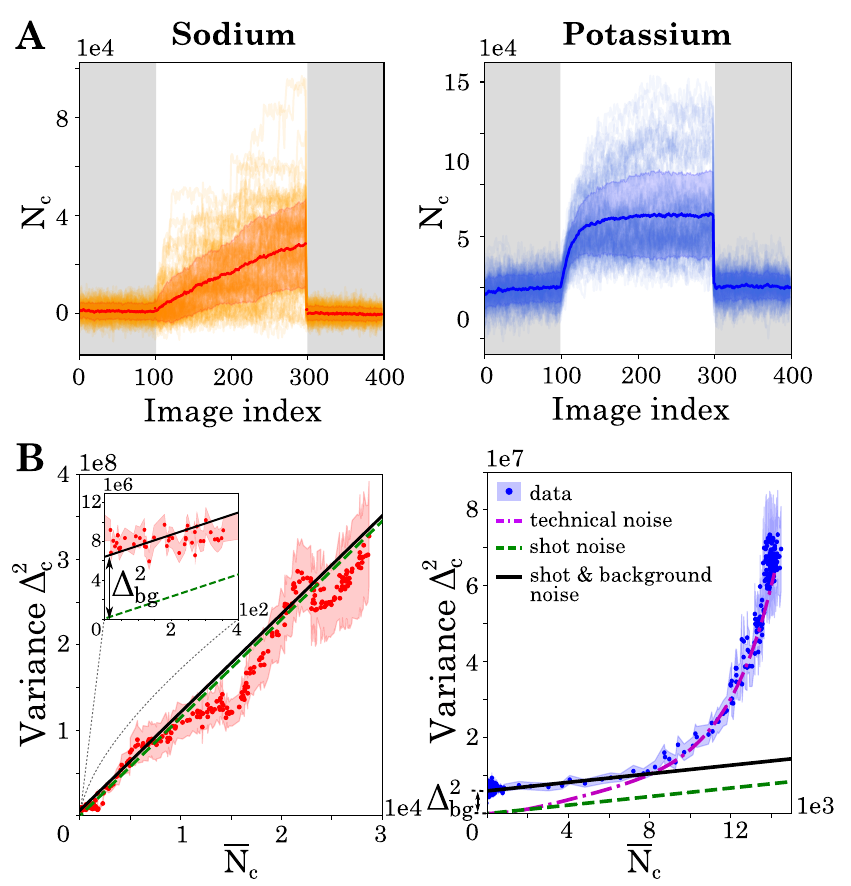}
	\caption{Characterization of atom number fluctuations for sodium (left) and potassium (right). \textbf{A:} Hundred time traces of sodium and potassium with mean and error band (shown as thick lines with shaded region around them). \textbf{B:} Dependence of variance on mean camera counts. For sodium (left) the inset shows the background noise level. \label{fig:VarianceAnalysis}}
\end{figure}

To extract expectation values through an ensemble average, we perform 100 repetitions of the previously described experimental sequence for sodium and potassium independently \footnote{Given the smaller size of the dataset, it was possible to increase the integration time $\tau$, as the heating of the MOT coils was less of a limiting factor. Therefore, we increased the integration time for sodium to \SI{200}{\milli\second} for the data shown in figure~\ref{fig:VarianceAnalysis} and \ref{fig:Na_AccurateCounting}.}. To further access the small atom regime for potassium in this analysis, we reduce the 2D-MOT power and do not perform pre-loading.
The observed dynamics are shown in figure \ref{fig:VarianceAnalysis}~A. We calculate the mean $\overline{N}_c$ and standard deviation $\stdNc$ of counts at each image index \cite{Muessel2013, Kang_2006}.
For the case of sodium, the dynamics is extremely slow and never reaches a stationary regime. Furthermore, the amplitude of the fluctuations is comparable to the average camera counts throughout the entire observation.
For potassium, the stationary situation is achieved on average after a few seconds of loading. Once again, we observe a strong dependence of the standard deviation on the average atom counts.

To study this dependence quantitatively, we trace the variance $\stdNc^2$ as a function of the average counts $\overline{N}_{c}$ in figure \ref{fig:VarianceAnalysis}~B.
For sodium the variance shows a linear dependence on the average counts with an intercept. This behavior can be understood by considering two independent noise sources. The first one is a background noise $\Delta_{\text{bg}}$, which is independent of the atom number and adds a constant offset to the variance. It originates from the readout noise of the camera and intensity-varying stray light. The second noise source is the atom shot noise, which describes the random variations due to the counting of atoms loaded until a given image index in the time trace.
Its variance is equal to the average atom number. 
The recorded camera signal is directly proportional to the atom number $N_{\text{c}} =C\,N_{\text{at}}$, leading through error propagation to a variance of $C\,\overline{N}_{\text{c}}$.
The two independent noise sources add up in their variances
\begin{equation}
\stdNc^2 = C \,\overline{N}_\text{c} + \Delta_\text{bg}^2\, . \label{eq:noisemodel}
\end{equation}

This theoretical prediction agrees well with the experimental observations. The calibration constant $C_{\text{Na}}= 1.15(5) \times 10^4$ and the background noise $\Delta_{\text{bg,Na}} = 2201(2)$ were independently extracted from a histogram plot, as described in section~\ref{Sec:Immersion}.
This validates our assumption that background and shot noise are the dominating noise sources for sodium.
Converting the camera counts back into atom numbers, we obtain a resolution of $0.20(1)\,$atoms, quantifying the quality of the observed single atom resolution \footnote{The connection of this resolution to the detection fidelity can be directly extracted from the histogram, shown in figure~\ref{fig:Na_AccurateCounting}.}.

For potassium, we observe a more complex behavior of the variance. In the regime of few counts the variance is again dominated by the background noise and the atom shot noise.
With the noise model \eqref{eq:noisemodel}, validated for sodium, we perform a fit to extract the calibration factor $C_{\text{K}}=560(140)$ and the background noise $\Delta_{\text{bg,K}} = 2450(140)$. The resulting atom resolution of $4.3(1.1)\,$atoms is similar to that achieved in precision experiments with Bose-Einstein condensates \cite{Strobel2014, Muessel2014}.

For higher atom numbers, we observe a non-linear dependence, which we attribute to technical fluctuations of the MOT. The MOT properties can be parameterized by the loading rate $\Gamma_{\text{load}}$ and loss rate $\Gamma_{\text{loss}}$.
Considering single atom load and loss only, they are connected to the atom dynamics through
\begin{equation}
N_{\text{at}}(t)=\frac{\Gamma_{\text{load}}}{\Gamma_{\text{loss}}}\big[1-\exp(-\Gamma_{\text{loss}}t) \big]\,.
\end{equation}
We fit each time trace with this solution and, hence, extract the distribution of $\Gamma_{\text{load}}$ and $\Gamma_{\text{loss}}$ across different runs. The variance in the atom dynamics, resulting from these fluctuations, is traced as the dash-dotted curve in figure \ref{fig:VarianceAnalysis}~B. In the high atom number regime it agrees well with our experimental observation. We expect to substantially reduce these fluctuations in the future by improving the stability of intensity, frequency, and magnetic field.

\section{Full counting statistics of sodium}\label{Sec:Immersion}
\begin{figure}[b]
	\includegraphics[width=1.0 \columnwidth]{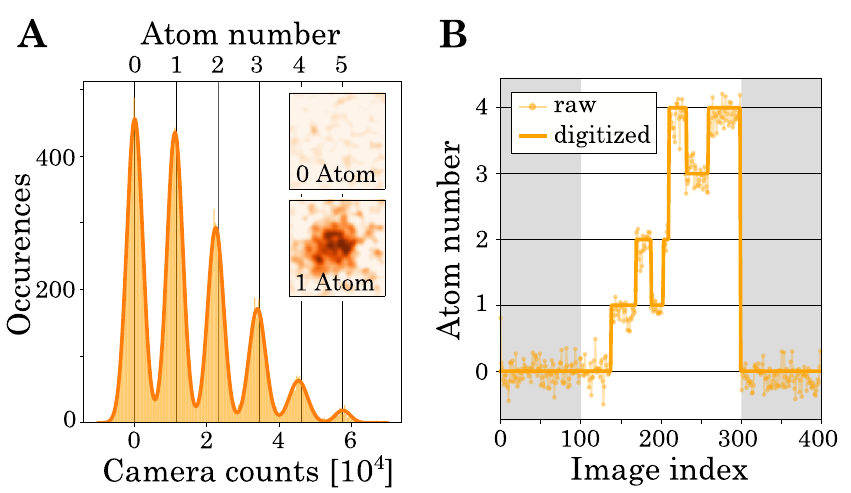}
	\caption{Accurate atom counting of sodium. \textbf{A:} Histogram of recorded camera counts. The calibration from camera counts to atom number is accomplished through Gaussian fits to distinct single atom peaks. Insets show average images of zero and one atom. \textbf{B:} Example time trace before and after digitization. \label{fig:Na_AccurateCounting}}
\end{figure}

Going one step beyond the statistical analysis of ensemble averages, we use the single atom resolution of sodium to extract its full counting statistics \cite{Widera_2020, Esposito_2009}. This requires the digitization of camera counts into discrete atom numbers \cite{Choi2007}, as presented in figure \ref{fig:Na_AccurateCounting}.
For this, we aggregate the camera counts of 100 runs into one histogram, which shows distinct atom number peaks.
The calibration from camera counts to atom counts is accomplished through Gaussian fits to individual single atom peaks. The distance between consecutive peaks corresponds to the calibration factor $C_{\text{Na}}=1.15(5)\times10^4$. The width of the zero atom signal sets the background noise limit $\Delta_{\text{bg,Na}} =2201(2)$ \footnote{These values are used in section~\ref{Sec:dynamics} in the analysis of the variance as a function of the mean camera counts $\overline{N}_\text{c}$.}. From the overlap of the peaks, we estimate the detection fidelity of atoms to $96(3)$\%.
With this calibration, we convert the time traces of camera counts into digitized atom count dynamics, as shown in figure \ref{fig:Na_AccurateCounting}~B.

\begin{figure}[t]
	\includegraphics[width=1.0 \columnwidth]{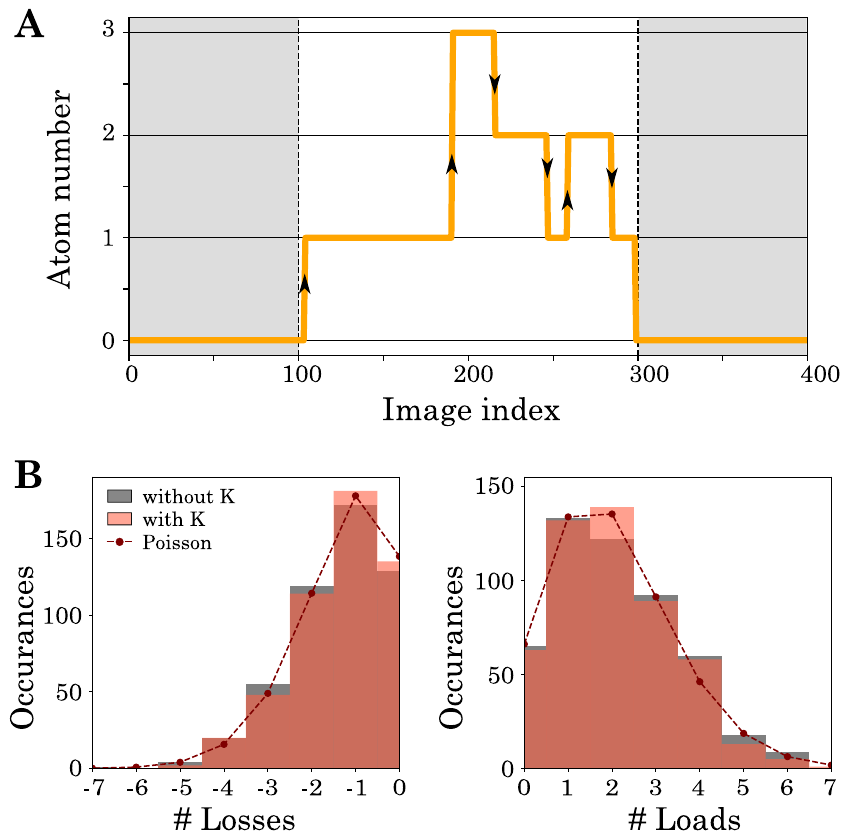}
	\caption{Counting statistics of sodium with and without potassium atoms present. \textbf{A:} Digitized example time trace of sodium with single atom load (loss) events marked with up (down) arrows. Only jumps during the MOT loading stage are taken into account. \textbf{B:} Histogram of the number of single atom losses and loads per time trace. The dashed lines show Poisson distributions with mean $\overline{N}_{\text{loss}}$ and $\overline{N}_{\text{load}}$ (extracted from the counting statistics). \label{fig:Na_CountingStatistics}}
\end{figure}

Each change in atom counts corresponds to a load or loss event with one or more atoms, as shown in figure \ref{fig:Na_CountingStatistics}~A. We observe that the dynamics are dominated by single atom events, as only $3$\% involve two or more atoms. Therefore, we neglect them in the following. We count the number of single atom events in each time trace and summarize them in a histogram, shown in figure \ref{fig:Na_CountingStatistics}~B. 
On average we observe $\overline{N}_{\text{load}} =2.02(6)$ loading events per time trace, which is much smaller than the total number of images $N_{\text{img}} = 200$ taken per time trace. Given that the atoms come from a large reservoir, namely the oven region, the loading rate is independent of the number of loaded atoms. From these observations, we describe the loading process statistically as a series of independent Bernoulli trials with a success probability $p_{\text{load}}$.
Therefore, the single atom loading probability is given by
\begin{equation}
p_{\text{load}}=\frac{\overline{N}_{\text{load}}}{N_{\text{img}}}\,.
\end{equation}
The large number of images and the low loading probability means that the number of loading events $N_{\text{load}}$ converges towards a Poisson distribution with mean $\overline{N}_{\text{load}}$. This stands in full agreement with the experimental observation.

Once an atom is present, it can be lost from the MOT with a probability $p_{\text{loss}}$. We observe an average number of $\overline{N}_{\text{loss}} =1.29(5)$ loss events per time trace. Since we do not distinguish between atoms, the number of atoms lost in each time step can be described by a binomial distribution. Therefore, the average number of single atom loss events per time trace $\overline{N}_{\text{loss}}$ enables us to extract the loss probability
\begin{equation}
p_{\text{loss}}=\frac{\overline{N}_{\text{loss}}}{\sum_{i} \overline{N}_{i}} \, .
\end{equation}
The normalization factor is the sum of average number of atoms present in each image $i$.
Similar to the loading case, we observe a Poisson distribution for the loss events with mean $\overline{N}_{\text{loss}}$, which can be attributed to the occurrence of only a few loss events over a large set of images.

\begin{table}
\caption{\label{tab:pload_ploss_results} Comparison of load and loss probabilities in a few atom sodium MOT with and without the presence of a potassium cloud. The uncertainties were obtained through bootstrap resampling.}
\begin{ruledtabular}
\begin{tabular}{lrr}
 & $p_{\text{load}}$ [\%] & $p_{\text{loss}}$ [\%]\\
\hline\vspace{-3 mm}\\
Without K & 1.06(3) & 2.76(23) \\
With K & 1.02(3) & 2.47(24) \\
\end{tabular}
\end{ruledtabular}
\end{table}
To study the influence of the large potassium cloud on the dynamics of the few sodium atoms, we compare the load and loss statistics of the sodium atom counts with and without potassium atoms present (see fig. \ref{fig:Na_CountingStatistics}~B). The extracted load and loss probabilities are summarized in table \ref{tab:pload_ploss_results}. The values corresponding to the absence and presence of potassium are indistinguishable to roughly within five percent. To exclude experimental errors, we repeated the analysis for various configurations of relative positions of the two clouds, magnetic field gradients and laser detunings. All results were compatible with our observation of no influence of potassium on the sodium atom dynamics.
We attribute these results to the extremely low density of the atomic clouds. To increase the density of both clouds in future studies, we plan to work at higher magnetic field gradients with water-cooled coils \cite{Roux2019}. At higher densities,  we expect to observe inter-species interaction, which should influence the loading dynamics similar to previous studies \cite{Castilho2019,Weber2010}.

\section{Outlook}\label{Sec:Outlook}
In this work, we presented a detailed experimental study of the stochastic dynamics of a few cold \textsuperscript{23}Na atoms immersed in a cloud of \textsuperscript{39}K atoms in a MOT. The experimental setup is designed to be directly extendable towards quantum degenerate gases through evaporative cooling \cite{Lamporesi2013a}. Defect-free optical tweezer arrays will provide high control over single atoms \cite{Barredo2016,Endres2016} and their repeated observation, as recently demonstrated for strontium atoms \cite{Covey2019}.

Our study opens the path for investigating time-resolved dynamics, including transport and thermalization, in atomic mixtures over a wide range of parameters. The emergence of ergodicity will become directly observable as the equivalence of the ensemble averages with time averages for individual time traces for sufficiently long times. Optimizing the photon detection should further allow us to reduce imaging times sufficiently to reach position-resolution \cite{Bergschneider2018, Bergschneider2019} without a pinning lattice \cite{Hohmann2017}. This will extend our work towards the quantum regime, in which we might continuously monitor thermalization of impurity atoms in a Bose-Einstein condensate \cite{Widera_2016, Mehboudi_2019}.

\section*{Acknowledgement}

The authors are grateful for fruitful discussions and experimental help from Apoorva Hegde, Andy Xia, Alexander Hesse, Helmut Strobel, Valentin Kasper, Lisa Ringena and all the members of SynQS. We thank Giacomo Lamporesi as well as Gretchen K. Campbell and her team for valuable input on the design of the experimental setup.

This work is part of and supported by the DFG Collaborative Research Centre “SFB 1225 (ISOQUANT)”. F. J. acknowledges the DFG support through the project FOR 2724, the Emmy- Noether grant (Project-ID 377616843) and support by the Bundesministerium für Wirtschaft und Energie through the project "EnerQuant" (Project- ID 03EI1025C).

\bibliography{bibliography}

\end{document}